\documentclass[aps,pra,groupedaddress,notitlepage,twocolumn]{revtex4-1}

\usepackage{amsmath}    % need for subequations
\usepackage{amsfonts}
\usepackage{bm}
\usepackage{amssymb}
\usepackage{graphicx}   % need for figures
\newcommand{\re}{\textrm{Re}}
\newcommand{\im}{\textrm{Im}}
\newcommand{\IN}{\textrm{in}}
\newcommand{\omegap}{\omega_{\textrm{p}}}
\newcommand{\bx}{\mathbf{x}}

\begin{document}

\title{Why the laser linewidth is so narrow: A modern perspective}

%\author{Alexander Cerjan and A.~Douglas Stone}
\author{Alexander Cerjan}
\author{A.~Douglas Stone}
\email[]{douglas.stone@yale.edu}
\affiliation{Department of Applied Physics, Yale University, New Haven, CT
06520, USA}

\date{\today}

\begin{abstract}
We review and interpret a modern approach to laser theory, steady-state \textit{ab initio} laser theory (SALT),
which treats lasing and amplification in a unified manner as a 
non-unitary scattering problem described by a non-linear scattering matrix. 
Within the semiclassical version of the theory the laser line has zero width as the
lasing mode corresponds to the existence of an eigenvector of the S-matrix with diverging eigenvalue due to 
the occurrence of a pole of the scattering matrix on the real axis.  In this approach the system is infinite from
the outset and no distinction is made between cavity modes and modes of the universe; lasing modes
exist both in the cavity and in the external region as solutions satisfying Sommerfeld radiation boundary
conditions.  We discuss how such solutions can be obtained by a limiting procedure in a finite box with damping according 
to the limiting absorption principle.  When the electromagnetic and matter fields are 
treated as operators, quantum fluctuations enter the relevant correlation functions and a finite linewidth is
obtained, via a generalization of SALT to include noise (N-SALT). N-SALT leads to an analytic formula for
the linewidth that is more general than all previous corrected versions of the Schawlow-Townes formula,
and can be evaluated simply from knowledge of the semiclassical SALT modes.  We derive a simpler
version of this formula which emphasizes that the noise is dominated by the fluctuations in the polarization
of the gain medium and is controlled by the rate of spontaneous emission.
\end{abstract}

\maketitle

\section{Introduction}

The laser has been a critical enabler of the modern discipline of quantum optics via its highly monochromatic and intense emission
properties.  The physical understanding of steady-state laser emission has presented challenges since the early days
after its invention, as first semiclassical \cite{haken_nonlinear_1963,lamb_theory_1964} and then fully quantum laser theories were developed \cite{lax_quantum_1966,scully_quantum_1967,graham_quantum_1968,graham_functional_1970}.  
Although these
theories were able to capture most of the relevant physics, they tended to incorporate some phenomenological 
or approximate steps due to the complexity of dealing with the full non-linear and space-dependent 
semiclassical or quantum laser equations. Over forty years ago, several pioneers of early laser theory, Lang, Scully and Lamb,
addressed
some of the weak points in the theory in a classic paper entitled,``Why is the Laser Line so Narrow?~A Theory of Single 
Quasi-mode Laser Operation'' \cite{lang_why_1973}.  This work was aimed at a fundamental challenge in laser theory, dealing with the
openness of the laser system, which invalidated standard Hermitian modal decompositions appropriate for closed systems.
Prior to this work essentially all theories were framed in terms of  ``quasi-modes'' of the type studied by Fox and Li and
were based on phenomenological modal equations with cavity damping inserted to represent the outcoupling of the modes \cite{fox_resonant_1961}.  The 
goal of Lang \textit{et al.}\ was to go beyond these quasi-mode equations for the 
laser field amplitudes to a more fundamental description which included both the cavity and exterior regions
explicitly (i.e.~a model of ``the universe''), and then work
back from the modes of the universe to the quasi-modes of the cavity region alone. 
Using this approach the authors (1) ``hope[d] to understand the mechanism leading to the extreme monochromaticity of the
laser radiation'', and (2) ``to investigate the sense in which the present calculation, which does not include the cavity dissipation
as a phenomological loss mechanism, still leads to and implies a fluctuation-dissipation theorem'' \cite{lang_why_1973}.  

While this approach to understanding the laser equations was groundbreaking at the time, there has been much progress in 
laser theory in the subsequent years and it seemed to us a good time to re-examine these questions from a modern
perspective.  One reason for this is that during the past decade 
a new theoretical framework has been developed, known as the steady-state \textit{ab initio} laser theory (SALT), which
treats the openness of an arbitrary cavity exactly within the semiclassical theory \cite{tureci06,tureci07,tureci08,ge10}, and thus
has provided new insights into lasing phenomena.  The resulting non-linear SALT equations 
for steady-state lasing can be solved efficiently using a non-Hermitian basis set method and have been shown to agree to very 
high accuracy with brute force simulations of the atomic lasing equations \cite{ge08,cerjan12}, including even N-level models with multiple 
independent lasing transitions \cite{cerjan_csalt_2015}.  The theory can also be generalized to treat injected signals (I-SALT), as will be relevant 
below \cite{cerjan_isalt_2014}.  Moreover, recently Pick \textit{et al.}\ have added Langevin noise to the SALT equations (N-SALT) and, using the SALT steady-state 
lasing modes as a starting point, calculated the quantum-limited laser linewidth analytically \cite{pick_linewidth_2015}.  Their analytic results 
have very recently been confirmed, again by direct simulations of the lasing equations, but with appropriate noise terms 
included \cite{cerjan_noise_2015}.  It should be said that many aspects of the results obtained through SALT and N-SALT were anticipated by prior 
theoretical approaches, a partial list of which includes the work by Spencer and Lamb \cite{lamb72}, Fu and Haken \cite{haken87,haken88,haken91}, and 
Henry \cite{henry86}; however SALT and N-SALT appear to be the most general, flexible and 
inclusive formalism yet developed for steady-state lasing. It treats the fields over the entire ``universe'' from the outset, 
but in a different manner from Lang \textit{et al.}, and does not arrive at modal equations with damping in the conventional sense.
We will now examine the questions raised by Lang \textit{et al.}\ from the 
perspective of SALT and N-SALT.

\section{Lasers as Scattering Systems}  
\subsection{Motivation and definition of the problem}
Most early lasers emitted from lasing modes which were concentrated along some optical axis, e.g.~the axis joining the 
centers of two mirrors or the geometric or index profile defining a waveguide or fiber, and so there was a tendency to 
develop laser theories focused on one-dimensional or axially symmetric equations, and this is overwhelmingly the type of 
formulation one finds in text books \cite{haken_LT,siegman,sargent}.  However in the past two decades a great variety of laser systems have been developed 
with complex two or three-dimensional cavities, such as microdisks and microspheres \cite{levi_microdisk_1992,mccall_1992,collot_microsphere_1993}, 
micro-toroids \cite{armani_toroid_2003}, deformed (chaotic) disk cavities \cite{gmachl98,lacey_directional_2003,redding_dcav_pnas_2015}, 
photonic crystal defect-based and surface emitting lasers \cite{painter_two-dimensional_1999,meier_laser_1999,imada_coherent_1999} 
and random lasers \cite{cao_random_1999,cao_microlaser_2000,polson_random_2001,bahoura_determination_2003,van_der_molen_spatial_2007,wiersma_light_2001,wu_random_2006}.  To describe such devices it was necessary to 
formulate laser theory in a manner that could both describe, and be implemented computationally, for cavities with an arbitrary geometry, 
and to include the full Maxwell wave equation, not its axial approximation.

\begin{figure}[t!]
\centering
\includegraphics[width=0.3\textwidth]{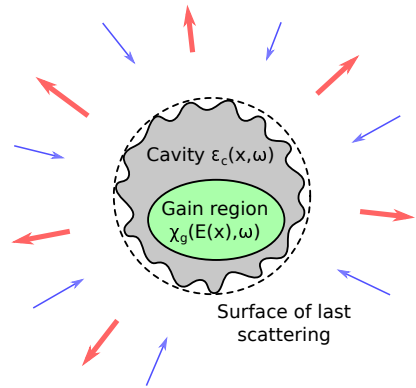}
\caption{Schematic of lasers as scattering systems:
a singly connected region surrounded by a last scattering surface 
containing an irregularly shaped cavity defined by a linear dielectric function $\varepsilon_c \ne 1$ and 
a non-linear gain region. Any incoming signals (blue arrows) are amplified and scattered by the cavity to outgoing states (red arrows);
lasing corresponds to outgoing solutions without any input, and is only possible at discrete frequencies for a given cavity
\label{fig:cavity}}
\end{figure}

Moreover, some of the new lasers (e.g.~random lasers) are extremely open systems, not even cavities in the usual sense; nor 
do they always emit well-behaved narrow beams of light but rather generate complex emission patterns in two and three dimensions.  
Thus it was crucial to find a description which did not over-simplify the spatial properties of the cavities and emission patterns, 
and which hence treated the cavity boundary and outcoupling exactly.
The solution to this challenge provided by SALT is to describe the laser as a scattering system.  Any laser consists of two essential elements: a cavity 
region, given as a linear dielectric function, $\varepsilon_c (\bx)$, which describes the scattering and confinement of photons, 
either injected externally or internally generated, and a gain region, containing the medium which will be pumped to provide 
optical gain (sometimes cavity and gain medium essentially coincide).
To include all cases the ``cavity'' will refer simply to a singly connected region large enough to enclose all the regions in which 
$\varepsilon_c (\bx) \neq 1$; and for simplicity we will assume the gain medium is fully contained within this region as depicted in Fig.~\ref{fig:cavity}.  

The laser so defined can function both as an amplifier of input radiation (when the gain medium is inverted) and as a self-organized 
oscillator/source of radiation in the absence of input, above the first threshold for lasing.  From this point of view the laser is a 
scattering system which both scatters and amplifies input radiation, and also, at discrete frequencies, scatters specific internally generated 
modes of outgoing radiation when sufficiently pumped.  However unlike linear scattering systems, the laser is subject to non-linear saturation 
effects which are fundamental to its stability and must be taken into account.  

Other non-linear effects within lasers, such as four-wave mixing, can destabilize their simple harmonic time-response, but these effects 
are neglected in SALT, and also (typically) in the earlier laser theories which motivated the work by Lang \textit{et al}.~\cite{lang_why_1973}.  Leaving out such effects is 
equivalent to assuming a stationary inversion density, giving rise to a gain susceptibility which does not mix frequencies.  This assumption 
is always valid for single-mode lasing and amplification, which will be the focus of this work; it is also valid under certain rather general conditions 
for multimode lasing in microlasers, and SALT has been shown to apply quantitatively to such cases \cite{ge08,cerjan12,cerjan_csalt_2015,cerjan_isalt_2014}.  

Before proceeding, it should be mentioned that the input/output theory developed by Collett and Gardiner (and others) 
also considers the cavity to be a scattering element \cite{collett_squeezing_1984,gardiner_input_1985} and is able
to treat the full quantum operator properties of the scattered fields.  This approach has certainly contributed
to our understanding of open quantum systems and the role played by the reservoirs and noise processes.
However, while there has been substantial
effort expended to describe lasing within this framework  \cite{glauber_quantum_1991,hackenbroich_field_2002,viviescas_field_2003,hackenbroich_quantum_2003},
it remains an open problem how to include the strong non-linearities present in active lasing systems in this
formalism; whereas at the semiclassical level this is done quite accurately by SALT.

\subsection{Scattering in the SALT formulation}

Initially we focus on the single-mode lasing case, as was done by Lang \textit{et al.}, so that the positive frequency component of the electric field will be given by:
\begin{equation} 
{\bf E}^{(+)}(\bx,t) = {\bf \Psi} (\bx) e^{-i\omega t},
\end{equation}
where the {\it real} number, $\omega$, is
either the input radiation frequency (for the amplifier) or the unknown frequency of the laser oscillator, which must be 
self-consistently determined.   ${\bf \Psi} (\bx) $ is the unknown spatial lasing (or amplified) mode, not assumed to be 
equal to any passive cavity mode or quasi-mode and is also self-consistently determined; such lasing/amplifier modes {\it describe the solution over all space}, not just inside the cavity.

Assuming stationary inversion and neglecting quantum and thermal fluctuations, we can describe ${\bf \Psi} (\bx)$ for an 
arbitrary laser by a non-linear wave equation of the form:
\begin{equation} 
[{\mathbf \nabla \times \nabla \times} + (\varepsilon_c +  4\pi \chi_g) k^2] {\bf \Psi} (\bx) = 0, \label{eq:maxWave}
\end{equation}
where $k=\omega/c$, and $\chi_g$ is the complex, frequency-dependent saturable gain susceptibility, which can be calculated 
for an arbitrary N-level atomic gain medium in steady-state \cite{cerjan12}.  For the case of a two-level gain medium it takes the form
\begin{equation}
\chi_g(\omega,\bx) = \frac{\frac{1}{4\pi}D_0 \gamma_\perp}{(i\gamma_\perp + (\omega - \omega_a))(1 + \Gamma(\omega-\omega_a)|{\bf \Psi} (\bx)|^2) }, \label{eq:chiDef}
\end{equation}
where $D_0$ is the pump parameter (proportional to the inversion), $\gamma_\perp$ is the dephasing rate of the polarization, 
$\omega_a$ is the atomic resonance frequency, and $\Gamma (\omega-\omega_a)$ is a Lorentzian centered at $\omega_a$.  $D_0$ and ${\bf \Psi}$ 
are measured here in appropriate dimensionless units \cite{ge08}, and this form assumes that there is only one mode, either lasing 
or amplified, active in the medium. (This is the only case we will consider explicitly here, but we will review the more general multimode SALT formalism in Appendix \ref{app:SALT}). The
non-linear gain susceptibility represents the polarization of the gain medium and is the source of the amplification and (above threshold) 
of laser emission.  The lack of any Langevin force terms in Eq.~(\ref{eq:maxWave}) is a result of the semiclassical approximation, in which the 
operators for the electromagnetic and polarization fields are replaced by their average values, eliminating the effects of quantum and thermal 
fluctuations.  Such terms are taken into account in N-SALT, which will be discussed briefly in Sec.~\ref{sec:noise}, but were 
neglected (intentionally) in the work of Lang \textit{et al.} 

Assuming the system is pumped below the first lasing threshold and the inputs are small enough to generate negligible saturation of the 
gain, this equation defines a linear non-unitary scattering (S-)matrix for input fields. Depending upon the geometry there are different 
natural basis states describing input and output modes, ${\bf H}_{\alpha}^{(-)}(\bx)$, ${\bf H}_{\alpha}^{(+)}(\bx)$, from and to infinity, i.e.~sufficiently far away from the cavity region,
\begin{align}
{\bf \Psi}(\bx) =& \sum_\alpha  [c_\alpha {\bf H}_{\alpha}^{(-)}(\bx) + d_\alpha {\bf H}_{\alpha}^{(+)}(\bx)]  \\
=& \sum_{\alpha, \beta}  [c_\alpha ({\bf H}_{\alpha}^{(-)}(\bx) + S_{\alpha,\beta} {\bf H}_{\beta}^{(+)}(\bx))]
\end{align}

If the pump is strong enough to invert the medium and create gain at the input frequency, the S-matrix will be amplifying, with 
eigenvalues greater than unity.  When functioning as an amplifier the boundary conditions are familiar: an input wave from infinity 
interacts with the cavity plus gain medium, scattering according to the appropriate dielectric or metallic boundary conditions in 
different regions of the cavity and giving rise to an outgoing wave which differs from passive scattering in that it is amplified 
and more photon flux emerges from the scattering region than enters it. The case of large inputs leading to a saturated gain and the 
resulting non-linear S-matrix can be handled by I-SALT and will be included in the discussion below.  

As noted, In this formulation there is no distinction between cavity modes and modes of the universe,
every amplified mode exists inside and outside the cavity as a steady-state scattering solution.
However we now consider to the more challenging and subtle case of laser oscillation in the scattering picture.

\section{Quasi-modes, Resonances and Quasi-bound States \label{sec:poles}}

Any scattering region of finite extent with a real potential or dielectric function, and hence with a unitary S-matrix,
will have associated with it resonant solutions; i.e.~stationary solutions at discrete 
complex frequencies which are purely outgoing.  It is easiest to think of these as evolving out of the bound
states of the system at real frequencies, $\tilde{\omega}_n$, which would occur if the region were surrounded by a perfect lossless 
reflecting barrier.  As the reflectivity of the fictitious barrier is reduced from unity to zero these real frequency solutions, 
corresponding to standing waves, evolve into complex frequency solutions with a substantial outgoing wave component (and no incoming 
wave from infinity). Due to the purely outgoing boundary conditions it can be shown that the imaginary part of the frequency must be negative, and
the S-matrix, continued to complex frequencies, will become infinite with simple poles at $\omega_n - i \gamma_n/2, (\gamma_n > 0)$.  
Hence these purely outgoing solutions are discrete and countably infinite, in one to one correspondence with the bound states of the closed cavity.

In the asymptotic region the frequency and wavevector of the solutions are linked by the dispersion relation and
the sign of $\gamma_n$ will cause the corresponding spatial solution to grow exponentially at infinity and carry a diverging outgoing 
photon flux. Hence these solutions are not physically realizable, but have physical significance because the S-matrix will have a resonant 
response near the real values $\omega_n$, with a bandwidth given by $\gamma_n$ (assuming well-separated high finesse resonances). 
The corresponding spatial solutions are referred to by various terms: quasi-modes, quasi-bound states and resonances. Historically the use of 
multiple terms seems to have arisen from the transition from 3D confinement, as e.g.~in a microwave resonator, where the connection to bound 
states is more apparent, to open, axial mirror-based resonators, where the resonances perpendicular to the mirror axis are essentially absent, 
and the Fox-Li method was developed to determine the high $Q$ resonances. However there is no fundamental difference between the various types 
of outgoing solutions: they all correspond to poles of the passive cavity S-matrix.

Such states also are very well-known from quantum scattering theory (where they are usually referred to as quasi-bound states), and are often 
considered to be physical states because if an initial condition is given corresponding to the quasi-bound state truncated to the ``cavity region'', 
it will leak out to infinity with a rate given by $\gamma_n$ (this property was explicitly demonstrated for the one-dimensional cavity treated by Lang \textit{et al.}).  
However in this context we would like to emphasize that the true quasi-bound states, including the exponentially growing component at infinity, 
are not physically realizable steady-state solutions and cannot correspond to the lasing mode.  
Also, since they occur at the poles of the S-matrix for the passive cavity, they can only 
describe the properties of the passive cavity, independent of the gain medium and spectrum.  Thus using ``quasi-mode'' also as a term 
for the lasing mode in the presence of gain can, and in our view often has, led to confusion.

\section{Threshold Lasing Modes and Constant Flux States \label{sec:tlm}}

\begin{figure}[t!]
\centering
\includegraphics[width=0.48\textwidth]{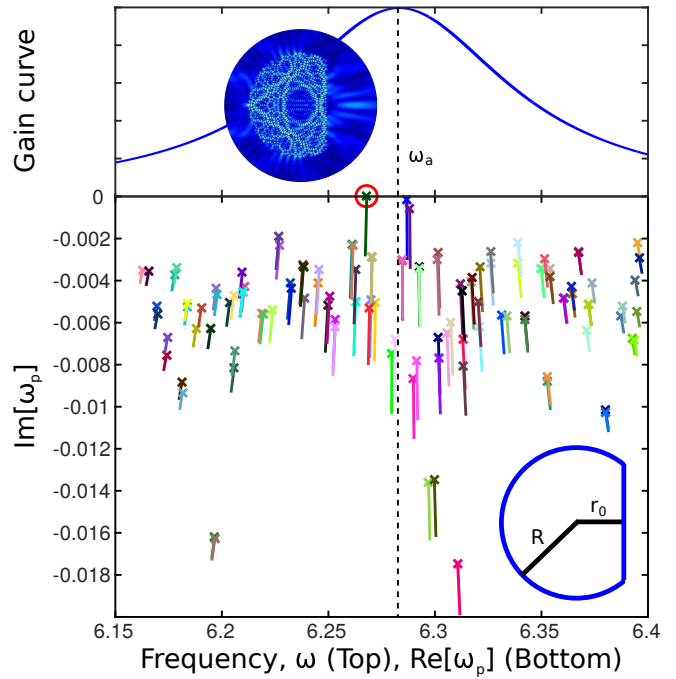}
\caption{Simulations for a D-shaped laser cavity with $R=4\mu m$, $r_0 = .5 R$, $n=3.5$ as shown in the
bottom schematic, with a uniform two-level atomic
gain medium characterized by $\lambda_a = 1\mu m$ and $\gamma_\perp = 10nm$. (Top) Plot of the Lorentzian gain profile
for the atomic medium. (Bottom) Plot of the motion of the poles of the S-matrix as the D-shaped cavity
is pumped to the first lasing threshold. The pole corresponding to the first lasing mode is seen to reach the
real axis (marked with a red circle), and the spatial profile of the lasing mode at threshold is shown
just above.
\label{fig:poles}}
\end{figure}

However, as described above, the properties of the laser/amplifier are not governed 
by a unitary S-matrix once the complex gain susceptibility is taken 
into account.  This complex susceptibility violates the Hermiticity of the wave equation even for the closed cavity and invalidates the
standard modal decomposition.  But it makes possible purely outgoing solutions at real frequencies, which hence are physical solutions 
over all of space.  The non-equilibrium steady-state gain susceptibility depends on the pump level, with its imaginary part changing sign 
from positive (absorbing) to negative (amplifying) at the transparency point, at which the upper and lower lasing levels are equally 
occupied and the susceptibility vanishes. At this pump value the poles of the cavity with gain coincide again with those of the passive 
cavity. As the pump is increased beyond the transparency point, all of the passive cavity poles continue to exist, but typically they 
move up in the complex plane towards the real axis \cite{ge10,chong12,pillay_2014}, at a rate that depends on their position in the gain spectrum. 
The 
S-matrix of the amplifier can be measured for real frequencies as the pump increases, and the height of the resonances corresponding 
to each pole increases (the transmitted flux is greater than the incident flux due to amplification), 
while their width decreases (gain narrowing), because the effective loss rate is proportional to the 
imaginary part of the pole frequency, which is decreasing.  If there is no input to the amplifier then there is still no emitted field 
and no indication within semiclassical theory of the approach to lasing (although when quantum fluctuations are included there will 
be enhanced noise, amplified spontaneous emission (ASE), near the gain center).  If lasing is achieved, one of the poles of the S-matrix reaches 
the real axis, gain completely balances loss, and we have a zero width resonance on the real axis.  
This process is demonstrated in Fig.~\ref{fig:poles} for a non-trivial two-dimensional cavity,
a chaotic ``D-shaped" dielectric cavity \cite{redding_dcav_pnas_2015}, where each pole is seen to move
towards the real axis an amount proportional its distance from the center of the gain curve and one of the poles
experiences enough gain to reach threshold. 

It first may seem that we are back at the familiar situation of true bound states, which correspond to poles on the real axis, but 
this is not so.  In the former case there is no photon gain and one simply has a decaying standing wave pattern, trapped in some region 
of space. Here we have a different type of object, which we call a ``threshold lasing mode'' (TLM).  
This type of solution never occurs in Hermitian theories, a purely outgoing solution at real frequency; it reflects the fundamentally 
non-Hermitian nature of the electromagnetic wave equations with a complex susceptibility, which is an effective theory in which the 
matter degrees of freedom are hidden.  Such a solution is characterized by a steady-state photon flux out to infinity, but with no 
unphysical growth of intensity {\it outside} the cavity and an emission pattern corresponding to the eigenvector of the S-matrix 
with a diverging eigenvalue. It is an example of a constant flux (CF) state: photons are generated within the cavity but are conserved 
outside the cavity, hence corresponding to a true physical lasing solution \cite{tureci06,tureci08,ge10}.  In Appendix \ref{app:SALT} we will show that each TLM is just one member of a 
complete biorthogonal basis set of CF states which forms a natural basis set for describing the non-linear lasing solution above threshold. 
%We will reserve the term lasing mode for this non-linear solution, which coincides with the TLM just at threshold.

However we still have a conceptual problem: the linear S-matrix for this amplifier is now infinite at the lasing frequency. 
Within linear theory any input would be infinitely amplified, violating conservation of energy.  
The resolution of this dilemma is
that infinitesimally above threshold the saturating non-linearity kicks in and stabilizes the solution.  In fact the steady-state lasing 
amplitude rises smoothly from zero and has no jump at threshold, just a discontinuous derivative.  Even though formally the S-matrix has 
a pole fixed on the real axis above threshold, any input wave at the lasing frequency simply increases the gain saturation and doesn't 
lead to a runaway of the output. Strictly speaking, any such input actually forces the pole off the real axis, as we will show with an 
example in Sec.~\ref{sec:example}.  Hence the non-linearity is essential for stablizing the linear CF solution.  However, the actual solution 
near threshold is accurately described by the single TLM calculated from the linear wave equation with gain, since the saturation term
is not strong enough to change the spatial form of the solution significantly.  

Returning to the fundamental question of modes of the universe vs. quasi-modes of the cavity, we summarize the SALT perspective.
\begin{itemize}  
\item  If one treats lasers as unbounded scattering systems from the outset, all solutions are defined both inside and outside the 
cavity region and there is no distinction between cavity modes and modes of the universe.
\item If the laser is functioning as an amplifier, the solutions are simply amplified scattering states determined by the boundary conditions
in the cavity region and the degree of gain at the input frequency.
\item If the laser is functioning as a self-organized oscillator the solutions are purely outgoing states corresponding to an 
eigenvector of the linear scattering matrix with diverging eigenvalue due to the existence of a pole on the real axis.  Such 
solutions require the presence of the saturating non-linearity to be physically stable.
\item For any given cavity and gain susceptibility such purely outgoing states can only occur at discrete frequencies because they 
arise from moving the discrete poles of the passive cavity onto the real axis.  Within semiclassical theory (neglecting 
quantum and thermal fluctuations of the fields) these solutions have precisely zero linewidth.  
\end{itemize}
We reserve the term ``lasing mode'' for these special discrete solutions. 
%and since these solutions exist 
%over all space, in this view lasing modes correspond to single ``modes
%of the universe''. Other, continuous in frequency, electromagnetic solutions will be called basis states or scattering states in the case of amplifed input states.

In summary then, our response to the question ``why is the laser linewidth so narrow?''~is that indeed this {\it is} due to the extreme limit of
gain narrowing: only at discrete frequencies can the linear gain diverge and generate self-organized oscillation.  As we 
will show in the next section, this zero-width resonance cannot be observed with an input signal, i.e.~in amplification mode.   
However these perfectly monochromatic purely outgoing solutions can be found by SALT and have been
verified by brute force FDTD simulations of the laser equations for many examples \cite{ge10,cerjan12,cerjan_csalt_2015}.
Such infinitely sharp lines are not of course obervable in true laser emission due to quantum 
fluctuations as we will discuss in Sec.~\ref{sec:noise}. However the smallness of these quantum effects lead to the extraordinary 
narrowness of the resulting lines, often many orders of magnitude narrower than the passive cavity resonance linewidths.

The effects of quantum fluctuations were intentionally not included in the work of Lang \textit{et al.}, who thus also found a delta function 
laser line within semiclassical theory.  In their approach they find that this extreme narrowing of the passive cavity 
resonance can be regarded as a ``mode-locking effect'' in which all of the modes of the universe within the passive cavity 
resonance linewidth contribute to one lasing quasi-mode with a single frequency. This point of view emerges from a limiting
procedure in which a finite system with quantized electromagnetic modes is studied and at the end the system size is 
taken to infinity in the presence of small fictitious absorption to damp out reflections from the boundaries as they tend to 
infinity.  The conditions for this procedure to reproduce the correct infinite system behavior are discussed in Sec.~\ref{sec:Modes} below.
We note that Lang \textit{et al.}\ expressed dissatisfaction with invoking gain narrowing to explain the laser linewidth, because 
in their view such arguments (at the time) did not take into account the non-linearity of the problem.
However the conceptual framework of SALT/I-SALT, outlined here, {\it does} take 
non-linearity into account, essentially exactly, for single-mode lasing. We will now illustrate the previous concepts with 
a concrete one-dimensional example, very similar to the one studied by Lang \textit{et al.}, in the process making all of our statements above mathematically precise.

%This makes clear that within semiclassical theory lasing only can occur at discrete frequencies and with zero linewidth.  All of these solutions are connected continuously to the discrete resonances of the passive cavity and in fact can arise in no other way; this is why lasing must occur at discrete frequencies and why the laser linewidth is so narrow (in fact infinitely narrow within the model considered by Lang et al.).

\section{Lasing modes as zero-width resonances: A simple example \label{sec:example}}

\begin{figure}[t!]
\centering
\includegraphics[width=0.3\textwidth]{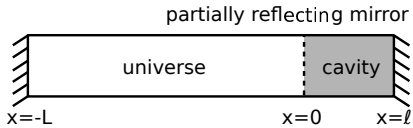}
\caption{Schematic showing the system studied by Lang \textit{et al.}\ \cite{lang_why_1973}, which
contains two regions, the cavity and the much larger ``universe'' (which also includes the cavity).  Damping
is added to the external region to prevent reflections from the boundary at $x=-L$, and as discussed below, this
must be done by a particular limiting procedure in order to obtain the correct result for the laser.
\label{fig:langCav}}
\end{figure}

The example studied by Lang \textit{et al.}\ was a one-dimensional cavity of length $L + l$, with perfectly reflecting mirrors
at $x=l,-L$ and a partially reflecting delta function mirror at $x=0$, as shown in Fig.~\ref{fig:langCav}. The ``cavity'' is the region $[0,l]$ and the ``universe'' is
the region, $[-L,0]$, where in treating lasing a small uniform absorption is added over all space, the limit $L \to \infty$ is taken,
and lasing emission is in a single direction towards $x =  - \infty$.
For such a system the passive cavity will have a reflection coefficient of modulus unity at all frequencies, 
and resonant reflection will be distinguished only by a long delay time in reflection and an associated build-up of field intensity in the cavity.  
We prefer to make a small modification of the Lang \textit{et al.}\ example and 
study a symmetric two sided partially transmitting (Fabry-Perot) cavity with laser emission to $x = \pm \infty$, and with passive cavity 
resonances distinguished by enhanced transmission, reaching unity at the center, and with width, $\gamma_c$. As with 
the Lang \textit{et al.}\ example, the electromagnetic part of the problem will be described by the scalar Helmoltz equation with gain.
However, unlike Lang \textit{et al.}, no limiting procedure is applied; our system is infinite from the start and no fictitious absorption is
added to avoid reflections from perfect mirrors far away from the cavity.  Hence, consistent with our scattering approach,
we start from continuous solutions for the 
passive cavity $+$ ``universe'', and will impose outgoing boundary conditions at infinity on the lasing solutions.

\begin{figure}[t!]
\centering
\includegraphics[width=0.48\textwidth]{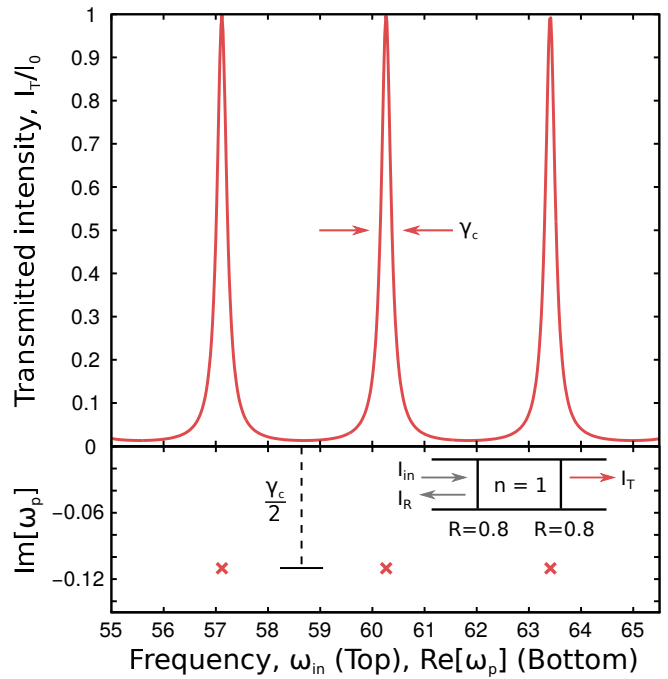}
\caption{(Top) Plot of the transmitted intensity of an injected signal with frequency $\omega_{\IN}$, 
into a passive, two-sided cavity with partially reflecting mirrors ($R=0.8$) on both ends, as
indicated in the schematic. (Bottom) Plot of the poles, $\omegap$ of the scattering matrix of the same
cavity, corresponding to purely outgoing solutions at complex frequency.
The real freuency resonances of the cavity, where complete transmission is seen,
are related to the poles of the cavity.
\label{fig:passive}}
\end{figure}

We begin by reviewing the properties of such a one-dimensional optical cavity. In Fig.~\ref{fig:passive}
we plot the transmitted intensity of an incident signal through the passive cavity at frequency $\omega_{\IN}$.
As expected
we observe a series of resonances of the cavity, spaced by the free spectral range, $\Delta = \pi/cL$,
with a width given by the passive cavity decay rate $\gamma_c = -c/(2L) \ln(R^2)$, where $L$ is the length of the cavity.
The incident and reflected waves on the left side of the cavity satisfy a continuity boundary condition,
with both incoming and outgoing components, whereas
the transmitted wave leaving the right side of the cavity also is continuous but in addition is purely outgoing.

Any purely outgoing monochromatic wave is described
mathematically by the Sommerfeld radiation condition \cite{sommerfeld},
\begin{equation}
\lim_{|x| \to \infty} |x|^{\frac{d-1}{2}} \left( \partial_{|x|} - i k\right) \psi(x) = 0, \label{eq:SRC}
\end{equation}
in which $\hat x = \vec x / |x|$, $\omega = c k$, and $d$ is the dimensionality of the system. 
In one dimension the external waves are simple plane waves (and not e.g.~cylindrical or spherical 
waves) and the Sommerfeld condition can be imposed anywhere outside the boundary;
thus at the boundary
on the right side of the cavity the field must be an outgoing plane wave,
satisfying,
\begin{equation}
\left[\left( \partial_{x} - i k\right) \psi(x)\right]_{x=L} = 0.
\end{equation}

As noted, lasing would correspond to a solution satisfying a purely outgoing boundary condition 
on {\it both} sides of the cavity (or in all directions in higher dimension).  If we attempt to impose this boundary condition
on the passive cavity we find that no solutions are possible at real frequencies, but there are discrete periodically
spaced solutions at 
\begin{equation}
\omega_m = \frac{\pi}{2cL} + m \Delta - i \gamma_c/2 \label{eq:QBeq}
\end{equation}
as shown in Fig.~\ref{fig:passive}.  The discreteness of 
the frequencies is a direct consequence of the Sommerfeld condition which can only be satisfied when there is constructive
interference in the cavity.  The fact that all the resonance frequencies have the same imaginary part is not generic, but is
particular to one-dimensional cavities of this type. As noted above, the presence of these purely outgoing (pole) solutions leads to
resonant transmission on the real axis centered at $\re[\omega_m]$ with width $\gamma_c$.  Applying the Sommerfeld condition 
to the outgoing solutions $e^{\pm ikx}$, and the relation $\omega = ck$ implies that $|\Psi (x)|^2 \sim e^{+\gamma_c |x|}$ at $\pm 
\infty$, the unphysical divergence we alluded to above.

%Lasing corresponds to the singular phenomena of having an outgoing signal from an optical
%cavity without an incident signal, and as such the wave function describing the lasing
%mode must satisfy the Sommerfeld radiation condition in all directions. This
%results in two important consequences. First, the Sommerfeld radiation condition yields a quantization
%condition on the allowable frequencies able to lase, resulting in
%a countably infinite set of potential lasing frequencies. This is in
%stark contrast to the case of transmission, where the frequency of the wave function is
%set by the incident signal, and thus allows for the full
%continuum of frequencies to be transmitted. 

In order to find normalizable physical solutions which are purely outgoing (i.e.~lasing solutions) we need to 
add the effect of the gain medium to the cavity dielectric function in the form of $\chi_g$ as defined in Eq.~(\ref{eq:chiDef})
above.  For this simple one dimensional scalar model (similar to Lang \textit{et al.}), below threshold, the electric field satisfies 
\begin{equation} 
[\partial_x^2 + (nk)^2 ] \Psi (x) = 0,
\end{equation} 
with the Sommerfeld boundary condition, 
where now the index of refraction $n = \sqrt{\varepsilon_c + 4\pi \chi_g} = n_1 - in_2$, and the gain non-linearity 
is neglected, because we are below threshold and these are not yet physically allowed solutions (they are simply modified active cavity resonances).
Note that the solutions inside the cavity here take the form $A \cos (n x) + B \sin (nx)$, i.e.~they are trigonometric functions 
of a {\it complex} argument.  This will remain true at threshold and is a key point which holds more generally: 
the electric fields for the physical system with 
gain cannot be purely real; if they were real they would conserve photon flux and describe true standing waves, whereas lasing 
or amplified modes must have running wave components. Any model which finds real spatial solutions for lasing or amplified modes has made 
an approximation. Such approximations have often been made in laser theory, including by Lang \textit{et al.}\
\cite{lang_why_1973,baseia_semiclassical_1984,penaforte_quantum_1984,dutra_spontaneous_1996} (see further discussion
in Sec.~\ref{sec:Modes}).

\begin{figure}[t!]
\centering
\includegraphics[width=0.48\textwidth]{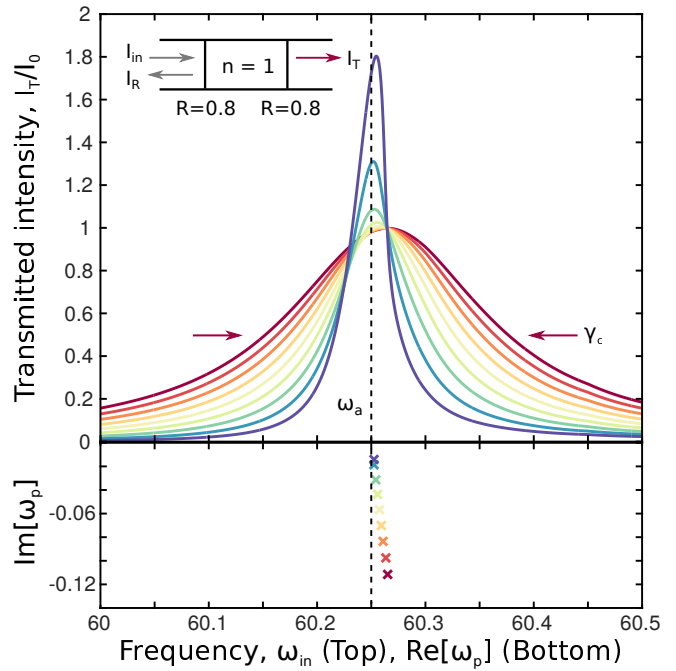}
\caption{(Top) Plot of the transmitted intensity of an injected signal as a function of frequency, $\omega_{\IN}$, illustrating line narrowing with
increased gain.  $I_{\IN} = I_0 = 0.01$, and the two-sided cavity has partially reflecting mirrors ($R=0.8$) at both ends, (see
schematic). The pump on the cavity is increased from the transparency point, $D_0 = 0$ (red),
to just above the first lasing threshold, $D_0 = 0.0040$ (blue), calculated using I-SALT \cite{cerjan_isalt_2014}. 
As can be seen, the width of the cavity
resonance narrows during this process. The slight shift of the resonance is due to line pulling from
central atomic transition frequency, $\omega_a = 60$, with $\gamma_\perp = 2$. At the transparency point this
system is identical to that shown in Fig.~\ref{fig:passive}, and the outermost red curve seen here is identical
to that from Fig.~\ref{fig:passive}. (Bottom) Plot of the
motion of the pole of the associated cavity resonance, $\omegap$, through the complex plane as the pump is increased. This is also calculated using
I-SALT, and thus includes the saturation of the gain medium due to the
injected signal, which prevents the pole from reaching the real axis even though the final pump
value is above the first non-interacting modal threshold. For the specific realization of the
motion of the pole shown, the injected field is taken at $\omega_{\IN} = 60.25$, which corresponds
to the peak in the blue curve. The pole is also seen to drift due to line pulling. Frequencies
and rates are reported in units of $c/L$.
\label{fig:active1}}
\end{figure}

When we add the gain medium to the cavity and pump above the transparency point
$(D_0 > 0)$ the imaginary part of $\chi_g$ corresponds to amplification.  This 
has an immediate effect, causing the poles to move upwards, as discussed earlier and as shown for this specific cavity in Fig.~\ref{fig:active1}.
While the pole solutions are still not physically realizable, we can observe a change
in the cavity properties via the behavior of the transmission resonances at real frequencies.  To calculate this effect accurately with
a finite input wave we must now use the full saturable gain susceptibility of Eq.~(\ref{eq:chiDef}).  The simulations
shown here are performed using I-SALT which provides a nearly exact description of
the steady-state single-frequency response of an active optical cavity to injected signals and any self-generated lasing
signals \cite{cerjan_isalt_2014}, including the spatial degrees of freedom. SALT and I-SALT are reviewed in more detail in Appendix~\ref{app:SALT}. 

%% An incident signal with frequency $\omega_{\IN}$ is injected into the active cavity with gain
%% as shown in the top panel of  Fig.~\ref{fig:active1}. As the pump is increased
%% from the transparency point, $D_0 = 0$ (red), which is identical to the passive
%% system studied in Fig.~\ref{fig:passive}, to larger values (bluer colors),
%% the injected waves with frequencies close to the cavity resonance experience greater amplification.
%% %The pole motion differs somewhat as the pump is increased depending on the input frequency, but the dominant 
%% %motion is always upwards. 
%% The upwards motion of the pole corresponding to the transmission resonance, $\omegap$, is plotted in the bottom panel of Fig.~\ref{fig:active1} for
%% an on-resonance incident signal with $\omega_{\IN} = 60.25$, changing in color as
%% $D_0$ is increased, with each color corresponding to the same value of the pump as used in the top panel. 
As the pole rises towards the real axis the amplified resonance narrows.  Due to the frequency-dependent saturation
of the gain medium, as already noted, injected signals with different frequencies will induce
somewhat different trajectories for the pole in order to maximize the effective gain
 \cite{cerjan_isalt_2014}. These are essentially line-pulling or pushing effects of the gain
medium, which for a single frequency are always towards the atomic transition frequency.  In the active cavity with dispersion,
the precise width of the transmission resonance can no longer be defined as the distance of the
pole from the real axis, as probing different frequencies will yield
slightly different distances of the pole from the real axis. These line-pulling effects lead to the mild
asymmetry in the series of curves shown in Fig.~\ref{fig:active1}.
%In some ways, this is akin to the measurement problem in quantum
%mechanics, the act of injecting a signal to observe the transmission response into
%an active optical cavity disturbs the properties of that cavity through gain-saturation,
%effecting the final result.  NOT SIMILAR ENOUGH, should omit.

The bluest (and narrowest) curve
shown in Fig.~\ref{fig:active1} corresponds to a pump value just {\it above} the non-interacting
first lasing threshold. This may seem initially puzzling; wasn't the lasing threshold the point at which
the pole reaches the real axis?  That definition only applies when there is no input signal.  With
an input signal, even at the lasing frequency, the saturation of the gain medium prevents the pole
from reaching the real axis, keeping the response finite, and the amplifed resonance retains a non-zero width.

\begin{figure}[t!]
\centering
\includegraphics[width=0.48\textwidth]{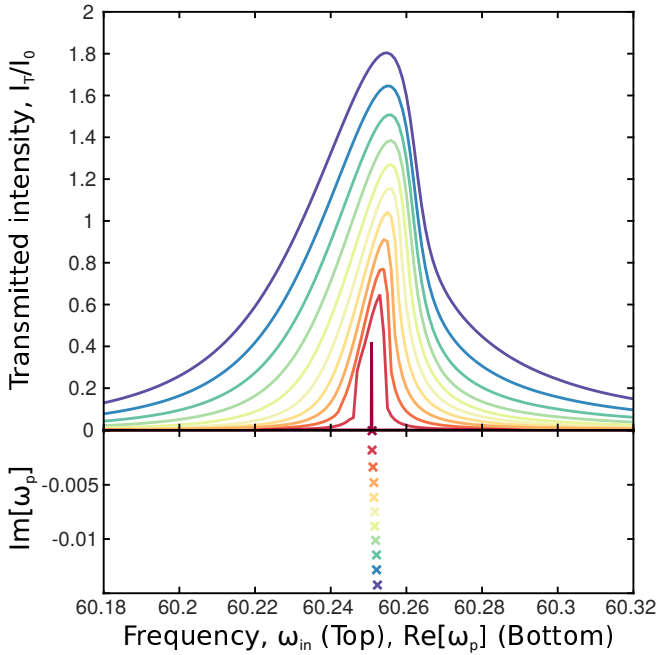}
\caption{(Top) Plot of the transmitted intensity of an injected signal as a function of frequency $\omega_{\IN}$,
into the active, two-sided cavity with partially reflecting mirrors ($R=0.8$) on both ends, as shown
in Fig.~\ref{fig:active1}. The pump on the cavity is held fixed just above the first lasing
threshold, $D_0 = 0.0040$, while the incident intensity is decreased from $I_{\IN}= I_0 =0.01$ (blue),
to $I_{\IN} = 0$ (red), where the cavity is operating as a free-running laser, calculated using I-SALT \cite{cerjan_isalt_2014}. As can be seen,
the width of the cavity resonance continues to narrow during this process, finally resulting
in the lasing signal with infinitesimal width. 
Note, the outermost blue curve shown here is the same as that shown in Fig.~\ref{fig:active1}.
(Bottom) Plot of the motion of the pole of the associated cavity resonance, $\omegap$,
through the complex plane as the injected signal is turned off for the specific realization of
$\omega_{\IN} = 60.25$. This is also calculated using
I-SALT, and thus includes the saturation of the gain medium due to the
injected signal, which prevents the pole from reaching the real axis until the injected
signal is turned off. For different values of $\omega_{\IN}$ (not shown) the pole is able
to reach the real axis before the injected signal is fully turned off.
\label{fig:active2}}
\end{figure}

To see the true zero-width semiclassical lasing solution we slowly turn {\it off} the incident signal
with the pump held above the
first lasing threshold.  With this protocol we see that the transmission resonance width 
continues to narrow,
as shown in Fig.~\ref{fig:active2}, while the pole continues
to move towards the real axis.
As the injected signal is turned off, the pole reaches
the real axis and the cavity begins to undergo self-organized lasing action,
shown as the reddest curve in the top panel of Fig.~\ref{fig:active2}.
This curve has precisely zero linewidth; no radiation comes out of the cavity except at this single frequency 
without input waves
(assuming effectively zero temperature in the cavity and neglecting quantum fluctuations until 
Sec.~\ref{sec:noise}).  This effect can never be fully achieved through gain-narrowing of an amplified input signal,
for the reasons we described above: gain saturation forces all poles off the real axis.  Hence this
effect does require correct treatment of the non-linearity as stated by Lang \textit{et al.}  However we have
shown here that it is possible to treat the non-linearity exactly at each stage and explain the 
zero width of the laser line quantitatively. 

The model we have treated here is essentially the same as that treated by Lang \textit{et al.}\ except
for the pedagogical difference of choosing a two-sided cavity in which one can study transmission resonances;
in the next section we will study exactly the model of Lang \textit{et al.}\ by a different but equivalent method.
As noted, in the standard SALT approach the system is infinite from the outset and there
is no dissipation outside the cavity, which is in our view the most natural model for a real laser system.
In contrast, Lang \textit{et al.}\ place their cavity in a much larger box which 
eventually will be taken to infinite size, and add dissipation external to the cavity, intended to damp unphysical reflections
from the boundaries of the larger fictional box.  We will discuss the effects of this difference in the next section.
%In our view the cavity determines the discrete poles, 
%and the cavity plus gain medium determine the lasing modes which are poles brought onto the real axis by
%the addition of gain, and stabilized to finite response by the saturating non-linearity.  There is no mode-locking
%of many modes of the universe at the lasing transition in this picture.  
%

\section{Lasing modes and the Limiting Absorption Principle. \label{sec:Modes}}

While it is clear that lasing modes should satisfy the purely outgoing (Sommerfeld) boundary condition,
this leads to a rather unfamiliar description of the spatial behavior of the lasing modes in terms of 
functions that are not power orthogonal, but instead satisfy a biorthogonality condition \cite{siegman,hamel_nonorthogonality_1989,hamel_observation_1990}.
Basis functions of this type, the CF states \cite{tureci06,tureci07,tureci08,ge10}, are central to the standard computational algorithm of SALT.
The approach taken by Lang \textit{et al.}\ of beginning with a finite system and adding damping to
remove the effect of the boundaries of the system is an alternative which
has evolved into the well-developed computational technique of surrounding a cavity or region of emitters with
a perfectly-matched (damping) layer (PML). When done correctly this can reproduce the imposition of Sommerfeld conditions
to arbitrary accuracy.  In fact, newer computational methods for solving the SALT equations \cite{esterhazy14} make use of PMLs and do not impose
the Sommerfeld boundary conditions in the straightforward manner of matching to outgoing solutions at a last scattering surface as did the 
earlier methods of satisfying the SALT equations
\cite{tureci08}. The PML approach gives essentially perfect agreement with the Sommerfeld approach to SALT.

However, as noted above, the correct lasing solutions {\it must} be complex functions in space, so as to 
represent the non-zero photon flux, which grows inside the cavity and remains constant outside.
For the simple 1D cavities studied here and by Lang \textit{et al.}, this means that the spatial solutions inside a constant
index cavity must be trigonometric functions of complex argument, corresponding to exponential growth towards
the mirrors as well as oscillation.  However in the work of Lang \textit{et al.}\ they find a lasing mode of the form
$E(x,t) \approx C_0e^{(\alpha - i\omega) t} \sin [k_0 (x-l)]$ for $ 0 < x <l$, where $\alpha$ is the difference between gain
and loss, $\omega$ is the lasing frequency, $C_0$ is a constant, and $k_0 = \Omega/c$ is the real wavevector 
corresponding to the passive cavity resonance frequency, $\Omega$.  Since the spatial part of this function is real, it
does not capture the photon flux through the partially transmitting mirror and is only an approximation to the complex
sine wave solution which is correct for this one-sided cavity.

In our view the origin of this problem is the neglect of the Limiting Absorption Principle \cite{ignatowsky_reflexion_1905,eidus_LAP_1962,schulenberger_limiting_1971},
which gives the correct mathematical procedure for recovering the Sommerfeld radiation condition from a finite system.  Lang \textit{et al.}\ solved the
spatial part of the problem for finite $L$, finding the quantized Hermitian basis states for the whole passive system (``modes
of the universe'') and then addressing the lasing problem within this Hermitian basis set.  Effectively the spatial 
part of the problem is fixed at finite $L$, and a small absorption, $\gamma$, is subsequently added everywhere in the universe so as to eliminate
the effect of reflection from the fictitious perfect mirror at $x=-L$.  For this procedure to work one needs to solve
the spatial problem {\it with} the finite damping, and carefully take the limit $L \to \infty$ prior to $\gamma \to 0$; this requirement
is known as the Limiting Absorption Principle (LAP). From a 
practical computational point of view the LAP means that one needs to solve the {\it non-Hermitian} 
spatial problem with Dirichlet (or other linear homogeous) boundary conditions {\it and}
finite damping, while maintaining the condition $\gamma k L \gg 1$.  The solutions so obtained have only exponentially small sensitivity
to the presence of the boundary and reproduce to high accuracy those obtained by actually imposing directly the Sommerfeld
conditions.  Hence they are significantly different from the Hermitian solutions calculated before the damping is added (or when the limits
are taken in the opposite order), which 
can carry no net flux as they are bound states.  The modern PML techniques are a means for introducing {\it non-uniform}
absorption which is impedance matched so that reflections are negligible even when $L$ is {\it not} very large and one 
reproduces the Sommerfeld solutions with a much smaller and manageable computational cell.  But the principle is basically
the same: in order to get the correct spatial solutions one must treat the non-Hermitian problem with damping from the 
outset.

To illustrate this explicitly we briefly analyze the Lang \textit{et al.}\ system from this point of view.
The system is the one-sided cavity between $[0,l]$ attached to the ``universe'' between $[-L,0]$, as shown in Fig.~\ref{fig:langCav}, with
perfect mirrors (Dirichlet boundary conditions), $E(l) = E(-L) = 0$,
where $E(x,t)$ is the electric field of the system. Before adding the gain and considering
lasing, we explicitly include the weakly absorbing material uniformly in the universe, and focus
on the region external to the cavity. The wave equation in this region is written as,
\begin{equation}
\left[ \partial_x^2 - \frac{1 + i \gamma}{c^2} \partial_t^2 \right] E(x,t) = 0, \;\;\; -L \le x < 0.
\end{equation}
We expand the field inside of the system in terms of a set of basis functions, 
$U_k(x)$, so that $E(x,t) = \sum_k U_k(x) e^{-i \omega t}$, and
solve for the form of these basis functions in the universe region as, 
\begin{multline}
U_k(x) = c_{1,k} e^{ikx - \frac{\gamma}{2}kx} + c_{2,k} e^{-ikx +\frac{\gamma}{2}kx}, \\ -L \le x \le 0, \label{eq:Uksol}
\end{multline}
where we have expanded the exponent for $\gamma \ll 1$.
The coefficients $c_{1,k}$ and $c_{2,k}$ and the quantized values of $k$ are set by the boundary conditions $U_k(-L)$, $U_k(l) = 0$,
and the matching conditions at $x=0$, and will vary with $L$, $l$ and the transparency of the mirror at the origin.
%The allowable values of $k = \omega/c$ are determined by the quantization condition
%at the interface between the cavity and the universe and depend upon the relative lengths
%of the cavity and the universe, as well as the strength of the partially reflecting mirror \cite{lang_why_1973}, but
%the exact details of this are not relevant here.

%The goal of this analysis is to reproduce the physical system we are familiar with, a cavity
%emitting into an infinite universe devoid of any absorption. Thus, there are two limits that
%must be taken on the system under consideration, the size of the universe must become infinite,
%$L \to \infty$, and the absorption must be decreased to zero, $\gamma \to 0$.
%The correct way to treat this problem is through the Limit Absorption Principle \cite{ignatowsky_reflexion_1905,eidus_LAP_1962,schulenberger_limiting_1971},
%the crux of which states that to recover solutions consistent with the Sommerfeld radiation condition, 

We first apply the LAP to obtain the Sommerfeld solution for this problem, taking the limit
$L \to \infty$ prior to $\gamma \to 0$.
Performing the limits in this order on the solution of $U_k(x)$, Eq.~(\ref{eq:Uksol}), one finds
that to satisfy the Dirichlet boundary condition, $U_k(-L) = 0$, we must choose $c_{1,k} = 0$,
as the term multiplying this coefficient is diverging as $-L \to -\infty$ for finite $\gamma$. Then, with the
spatial limit taken, we can safely reduce the absorption in the problem to zero,
recovering
\begin{equation}
U_k(x) = c_{2,k} e^{-ikx}, \;\;\;\; -\infty \le x \le 0,
\end{equation}
finding that all of the solutions of our problem are purely outgoing from the cavity,
consistent with the Sommerfeld radiation condition, Eq.~(\ref{eq:SRC}). However, if these two limits are taken in the opposite order,
setting the absorption to zero before letting the length increase to infinity,
we find that
\begin{equation}
\tilde U_k(x) = b_k \sin(k(x+L)), \;\;\;\; -L \le x \le 0, \label{eq:Ukwrong}
\end{equation}
which is exactly the result of Lang \textit{et al.}\ (outside the cavity region).  With this result the matching to the cavity allows for real sine wave solutions,
whereas with the outgoing solutions only complex sine waves are allowed in the cavity.  The complex sines provide the outgoing
flux at the partially transmitting mirror which is needed for continuity with the field outside the cavity.

\begin{figure}[t!]
\centering
\includegraphics[width=0.45\textwidth]{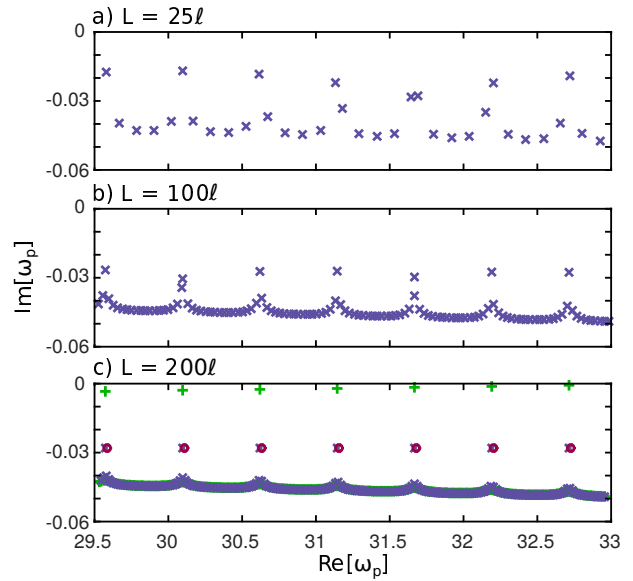}
\caption{Plot of the resonances of the finite system with external damping depicted in Fig.~\ref{fig:langCav} for three
different values of $L$, $L=25l$, $L=100l$, and $L=200l$, from top to bottom, except that the optical confinement
is instead provided by a passive cavity dielectric of $n = 6$ in the cavity, rather than a partially
reflecting mirror. In the external region an absorption of $\gamma = 0.003$ is present. In (c) the isolated solutions of the
problem with damping are compared to the resonances
of the same cavity subject to the Sommerfeld boundary condition on an infinite system without damping, given by Eq.~(\ref{eq:QBeq})
(red circles). Perfect agreement is found as discussed in the text.
(Note that due to the presence of the passive cavity dielectric constant 
the right side of Eq.~(\ref{eq:QBeq}) must be multiplied by $1/n$ in making the comparison). 
The green crosses shown in (c) show the
locations of the resonances when uniform gain is added to the cavity to find the threshold lasing mode, $n = 6 - .005i$. 
As can be seen, the poles corresponding to resonances of the cavity have moved towards the real axis,with the rightmost
one reaching the lasing threshold, while those corresponding to the resonances of the universe have stayed relatively fixed,
as discussed in the text.  This is thus an alternative method to implement the Sommerfeld conditions for lasing modes.
\label{fig:langFig}}
\end{figure}

\begin{figure}[t!]
\centering
\includegraphics[width=0.4\textwidth]{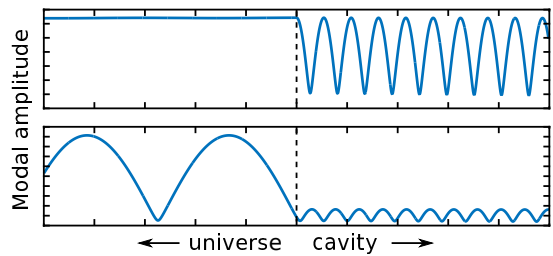}
\caption{Plot of the modal amplitudes for a small region near the cavity-universe boundary for two modes,
one corresponding to a resonance of the cavity (top), and the other corresponding to a resonance of the external universe (bottom).
As can be seen, the cavity mode is a nearly pure running wave after it exits the cavity.
\label{fig:langFig2}}
\end{figure}

These conclusions can be further reinforced by directly solving for the eigenfunctions of
the finite system studied by Lang \textit{et al.}, shown in the three panels of Fig.~\ref{fig:langFig}. 
As $L$ is increased, the modes of this finite system separate into
two sets. One set consists of discrete isolated solutions with smaller imaginary frequency (damping); the number of such solutions
does not change with increasing $L$. The
second set has larger damping, and gets denser and denser, forming a psuedo-continuum as $L$ is increased.  
The first set corresponds
precisely to the complex resonance frequencies of the passive cavity, as 
can be confirmed by plotting the known resonances of the cavity given by Eq.~(\ref{eq:QBeq}),
shown in Fig.~\ref{fig:langFig}(c).  The spatial functions corresponding to the passive resonances are identical to the
resonant solutions of the infinite system without damping {\it inside the cavity}, as there is no loss there, and
the boundary condition at the mirror is the same.  Outside the mirror these solutions experience a damping not
present in the infinite system and hence decay slowly as they propagate outwards.  On the scale shown in
the top panel of Fig.~\ref{fig:langFig2} one sees only that the outgoing boundary condition is satisfield and not the
weak decay. In contrast, the spatial solutions corresponding to the dense set of damped modes of the universe
can be seen in the bottom panel of Fig.~\ref{fig:langFig2}; they form standing waves outside the cavity, with small
leakage into the cavity, and do not satisfy the purely outgoing conditions. Hence simply discarding the latter and
keeping the first set is equivalent to finding the Sommerfeld solutions of the passive problem without damping, in the 
vicinity of the cavity.

Finally, we can make contact with the SALT approach by adding gain (taken to be uniform for convenience) in the cavity region
while leaving the damping present outside the cavity.  The added gain has a much bigger effect on the cavity resonance solutions
which have high amplitude there, than on the external standing wave solutions.  For large enough gain 
one of the cavity resonances can reach the real axis and will then correspond to a 
threshold lasing mode for uniform gain, as shown in Fig.~\ref{fig:langFig}c (green crosses).
This is the essence of the method adopted in Ref.~\cite{esterhazy14}
except in that work a PML damping layer is used for higher computational efficiency and the saturating non-linearity is
taken into account. Thus the Lang \textit{et al.}\ method agrees with
the SALT approach if care is taken to apply the limits in the right order.

%As gain is added to the cavity region, all of the resonances will move towards the real
%axis, analogous to the behavior seen in Fig.~\ref{fig:poles}, however, at the first lasing
%threshold, only a single resonance will reach the real axis, as discussed in Sec.~\ref{sec:example},
%and this resonance will satisfy a purely outgoing boundary condition.

%
%Furthermore, simply re-inserting the absorption by hand, such that $\psi(x,t) = \sum_k \tilde U_k(x) e^{-i\omega t - \gamma t}$,
%does not rectify the situation. Even though these solutions will decay in time,
%the structure of $\tilde U_k(x)$ in the universe region of the system, Eq.~(\ref{eq:Ukwrong}),
%means that the cavity is being subjected to incoming signals, which does not
%represent how lasers operate.

%Finally, to make contact with numerical methods as discussed above, if one solves this
%simple model outside the cavity for finite $L$ and $\gamma$ one finds
%%we can use the general solution of the modes in the universe, Eq.~(\ref{eq:genSolUk}),
%%to understand solutions found by numerical methods which truncate the simulation region.
%%Using the Dirichlet boundary condition, $U_k(-L) = 0$, alongside a finite absorption,
%%we find that
%%% \begin{align}
%%% 0 = & c_{1,k} e^{-ikL + \frac{\gamma}{2}kL} + c_{2,k} e^{ikL -\frac{\gamma}{2}kL}, \notag \\
%%% c_{1,k} = & -c_{2,k} e^{2ikL} e^{-\gamma kL}.
%%% \end{align}
%\begin{equation}
%c_{1,k} = -c_{2,k} e^{2ikL} e^{-\gamma kL},
%\end{equation}
%independent of the value of $c_{2,k}$.
%As long as $\gamma k L \gg 1$ this implies that the spurious incoming wave contribution
%to the solutions will be exponentially suppressed, yielding a good approximation to the
%correct Sommerfeld boundary conditions.

\section{Quantum limited laser linewidth: an alternative form \label{sec:noise}}

Since this paper is devoted to the ``SALT perspective'' on the laser linewidth, we include here
a section on the finite laser linewidth due to quantum fluctuations, which has recently been 
addressed within two different generalizations of SALT, one based on input-output theory \cite{chong12,pillay_2014}
and another based on non-linear perturbation theory around the SALT solution \cite{pick_linewidth_2015}.  
The fundamental origin of a finite quantum linewidth for the laser is the presence of spontaneous emission,
which leads to random jumps in the time-dependent phase of the output field.  The resulting phase diffusion is linear in
time and the phase diffusion coefficient determines the intrinsic linewidth. The basic physical understanding of this
goes back to the 
seminal work of Schawlow-Townes \cite{schawlow_1958}, which was subsequently substantially extended by Lax, Henry, Petermann and many others.  

Traditionally, the Schawlow-Townes (ST) linewidth is written in terms of the cavity
decay rate, $\gamma_c$, as
\begin{equation}
\delta \omega_{ST} = \frac{\hbar \omega_0 \gamma_c^2}{2 P}n_{sp}, \label{eq:STlw}
\end{equation}
where $\omega_0$ is the lasing frequency and $P$ is the total output power.  Here we have chosen to include
 in the ST formula the ``incomplete'' inversion factor, $n_{sp}$, which is sometimes treated as a correction to it
($n_{sp} \geq 1$ measures the degree of inversion of the laser frequency and tends to unity when the lasing transition is completely inverted, 
the limit assumed by Schawlow and Townes in their original work) \cite{lax_conf_1966}.

As discussed extensively above, semiclassical laser theory neglects quantum effects such as spontaneous emission
and leads to a delta function linewidth.  
In order to go beyond this one needs to treat all three fields, the electromagnetic field,
the polarization, and the inversion as quantum operators and consider their higher moments.  The laser linewidth will be 
obtained from the correlation function of the electric field near the lasing frequency. It is well known 
that the effects of these quantum fluctuations can be captured by adding appropriate Langevin (random) noise terms
to the equations of motion for the three fields \cite{haken}.  These noise terms will be related to the dissipation and
amplification in the system by the appropriate fluctuation-dissipation relations.  This was one of the questions raised
in the original work by Lang \textit{et al.}  It is now understood that semiclassical theory violates these relations but they
are restored by including the Langevin terms. The recent work by Pick \textit{et al.}\ \cite{pick_linewidth_2015} provides the most
complete treatment of these noise terms and the formalism they develop is referrred to as N-SALT (SALT + noise).  

Pick \textit{et al.}\ begin with the time-dependent wave equation for the field and add the appropriate noise term 
\begin{equation} 
[{\bf \nabla \times \nabla \times  E} (\bx,t)  + \varepsilon_c \ddot{\mathbf{E}}(\bx,t)] = 4 \pi \ddot{\mathbf{P}} +  \tilde{\mathbf{F}}_{tot} ,
\end{equation}
in which $\tilde{\mathbf{F}}_{tot} (\bx,t) $ is the total effective noise driving the field in the cavity in units of electric field.  Since this random drive is broadband, it is no
longer possible to assume a monochromatic steady-state exists as we do in semiclassical theory.  However only the noise 
near the lasing frequency is strongly amplified and by expanding the field around the SALT steady-state solution one can derive
non-linear coupled mode type equations which permit \textit{ab initio} calculation of the relevant correlation functions and
hence of the most general form of the laser linewidth yet derived \cite{pick_linewidth_2015}.  This form includes all of the important 
corrections to the ST formula: the correction due to relaxation oscillations ($\alpha$ factor) \cite{henry82,henry86}, due to openness of the cavity
(Petermann) factor \cite{petermann_calculated_1979,haus_excess_1985,siegman_excess_1989,hamel_nonorthogonality_1989,hamel_observation_1990}, 
due to dispersion (bad-cavity factor) \cite{lax_conf_1966,kolobov_role_1993,kuppens_quantum-limited_1994,vanexter_theory_1995,kuppens_evidence_1995}, and corrections not previously studied due to the spatial 
variation of $n_{sp}$ \cite{lax_conf_1966}.  We will not describe this formalism in any more detail here, but will briefly reproduce the essential
result at the end of this section.

However, in this section we would like to make a different conceptual point, not discussed in the earlier papers, relating to the 
interpretation of the Schawlow-Townes formula.  It seems initially puzzling that the noise due to spontaneous 
emission in the gain medium should lead to the standard ST formula, which is expressed only in terms of the
passive cavity linewidth, $\gamma_c$, the total output power, and the lasing frequency.  One would expect
the spontaneous emission rate to appear as a fundamental quantity, not the cavity decay rate; and here we will show that the ST linewidth can
(and perhaps should) be written in an equivalent form where this is the case.

%However, this is misleading, as the physical source of the noise arises
%from spontaneous emission of the gain atoms, and should be characterized in terms
%of the parameters of the gain medium such as the polarization dephasing rate, $\gamma_\perp$,
%and not from the fluctuations in the field from photons outside of the cavity,
%which are mediated by the boundary of the cavity, $\gamma_c$. This oddity has persisted through the
%discussion of the laser linewidth because at threshold, the gain from the
%material must balance the losses of the cavity, and thus the material implicitly knows
%about the cavity decay rate through being pumped above threshold. 

%with $\tilde{\mathbf{F}}_{tot}(\bx,t) = i\sqrt{2\pi \hbar \omega} F_{tot}(t) \mathbf{\Psi}_0(\bx)$ maybe???

The total Langevin force, $F_{tot} (t)$, in units of the quantized field amplitude operator, $\tilde{\mathbf{F}}_{tot} = i\sqrt{2\pi \hbar \omega} F_{tot}(t) \mathbf{\Psi}_0(\bx)$, 
has a direct contribution from the amplitude fluctuations of the field, $F_a (t)$, due to quantum
fluctuations from outside the cavity, but has also has
indirect contributions from Langevin forces driving the atomic polarization and inversion, $F_{p}$ and $F_d$ respectively \cite{haken},
\begin{equation}
F_{tot}(t) = F_{a}(t) + \left(\frac{1}{\omega_0 - \omega_a + i\gamma_\perp}\right) \sum_\alpha g_\alpha F_{p}(t)
\end{equation}
where the summation runs over all of the atoms in the cavity at locations $x_\alpha$ and 
\begin{equation}
g_\alpha = i \boldsymbol{\theta}\cdot \mathbf{\Psi}_0(\mathbf{x}_\alpha) \sqrt{\frac{2\pi \omega_0}{\hbar}},
\end{equation}
is the dipole coupling element for the quantized electric field.
$\boldsymbol{\theta}$ is the dipole matrix element for the lasing transition, and $\mathbf{\Psi}_0(x)$ is the normalized mode profile, $\int |\mathbf{\Psi}_0|^2 dx = 1$, of the electric field at frequency $\omega_0$.
The correlation function for the total effective Langevin force can then be calculated \cite{haken} as,
\begin{multline}
\langle F_{tot}^\dagger(t) F_{tot}(t')\rangle = \delta(t-t') \bigg(\gamma_c n_{th} \\
 +  \sum_\alpha \frac{|g_\alpha|^2}{\gamma_\perp^2} \left[\gamma_\perp(1 + \langle d_\alpha \rangle) + \frac{\gamma_\parallel}{2}(d_0 - \langle d_\alpha \rangle)\right]\bigg), \label{eq:Ftot}
\end{multline}
in which $n_{th}$ is the number of thermal photons at the lasing frequency, $\gamma_\perp$
is the dephasing rate of the polarization, 
and $d_\alpha$ is the inversion of the $\alpha$th atom, which would relax to the steady-state
value of $d_0$ in the absence of an electric field at the rate $\gamma_\parallel$. 
We have also simplified by assuming the lasing signal to be on resonance with the atomic transition, 
$\omega_0 = \omega_a$, and that the dipole matrix element and field are co-linear, $\boldsymbol{\theta}\cdot \mathbf{\Psi}_0 = \theta \Psi_0$. 

The first term in Eq.~(\ref{eq:Ftot}) corresponds
to the contribution from thermal and zero point fluctuations from outside the cavity to the field inside the cavity;
as such it is controlled by the decay rate of the cavity, $\gamma_c$.  The second and third terms 
stem from spontaneous emission events leading to the dephasing of the atomic polarization, and these terms
depend on the parameters of the gain medium. For optical frequencies, the number of thermal photons
present is negligible, so the direct field fluctuations do not have any effect on resulting linewidth.
Moreover, for most lasers (class A and B) \cite{ohtsubo}, $\gamma_\parallel \ll \gamma_\perp$,
so that only the second term in Eq.~(\ref{eq:Ftot}) remains,
\begin{equation}
\langle F_{tot}^\dagger(t) F_{tot}(t')\rangle = \delta(t-t') 
\frac{1}{\gamma_\perp} \sum_\alpha |g_\alpha|^2 (1 + \langle d_\alpha \rangle).
\end{equation}
Assuming that the atoms are uniformly distributed over the cavity,
the sum over the atoms in the cavity can be converted into an integral,
\begin{multline}
\langle F_{tot}^\dagger(t) F_{tot}(t')\rangle = \delta(t-t') 
\frac{2|g'|^2}{\gamma_\perp} \\ \times \int \left(\frac{N_2(x)}{N_2(x) - N_1(x)}\right) D(x) |\Psi_0(x)|^2 dx, \label{eq:Ftot2}
\end{multline}
where $N_2(x)$ ($N_1(x)$) is the density of atoms in the upper (lower) lasing level,
and $g_\alpha = g' \Psi_0(x_\alpha)$.  Equation~(\ref{eq:Ftot2}) now expresses the Langevin force in terms of 
the incomplete inversion factor \cite{lax_conf_1966},
 \begin{equation}
n_{sp}(x) = \frac{N_2(x)}{N_2(x) - N_1(x)}. \label{eq:incomp}
\end{equation}

Traditionally at this step the atomic
populations are approximated to be uniform in space which clamps the inversion
to its value at the lasing threshold, $D(x) \approx D_0 = D_0^{(thr)}$ \cite{haken}. 
In this approximation, the integral in Eq.~(\ref{eq:Ftot2}) becomes just $D_0^{(thr)} n_{sp}$, and 
can be related back to the cavity decay rate by
\begin{equation}
\frac{D_0^{(thr)} |g'|^2}{\gamma_\perp} = \frac{\gamma_c}{2}, \label{eq:gcApprox}
\end{equation}
which states that at the lasing threshold, gain must equal loss.
Using this approximation contributes one factor of $\gamma_c$ in the standard ST linewidth
formula, Eq.~(\ref{eq:STlw}). However this is a rather crude approximation.  Due to spatial hole-burning the atomic populations generally 
vary substantially in space, and as a consequence the overall inversion is not clamped at the threshold
value.  

%The N-SALT treatment takes these effects into account exactly and hence should be more
%accurate than standard treatments.  Moreover, as mentioned above, using this relationship obscures the fact that the physical
Using Eq.~(\ref{eq:gcApprox}) obscures the fact that the physical
origin of the noise is spontaneous emission in the gain medium. It then is more natural to 
relate the linewidth to the rate of {\it spontaneous emission}, $\gamma_{sp}$, which can be calculated
from the rate of stimulated emission, $\gamma_{SE}$ \cite{cerjan_csalt_2015}, as
\begin{equation}
\gamma_{sp}(x) = \frac{\gamma_{SE}(x)}{2 n_p} = \frac{2 |g'|^2}{\gamma_\perp} |\Psi_0(x)|^2,
\end{equation}
where $n_p$ is the number of photons in the steady-state lasing field, and is related to the intensity by $I_0 = 2\pi \hbar \omega_0 n_p$.
Thus, we can rewrite Eq.~(\ref{eq:Ftot2}) in terms of the rate of spontaneous emission, keeping the full space
dependence of the problem, as
\begin{equation}
\langle F_{tot}^\dagger(t) F_{tot}(t')\rangle = \delta(t-t') 
\int \gamma_{sp}(x) D(x) n_{sp}(x) dx. \label{eq:Ftot3}
\end{equation}

As noted above, the coefficient of the Langevin noise determines the laser linewidth by
standard manipulations.  Using the more correct form of this coefficient above leads to
a quantum-limited linewidth,
\begin{equation}
\delta \omega = \frac{\int \gamma_{sp}(x) D(x) n_{sp}(x) dx}{2 n_p}.
\end{equation}
Typically, the next step is to relate this to the total output power of the cavity via
\begin{equation}
P_{ST} = \hbar \omega \gamma_c n_p,
\end{equation}
which is also only an approximate relation, neglecting spatial variations of the intensity.
Using both of the approximations mentioned above and some other simplifications valid
for high $Q$ cavities (and neglecting the amplitude contribution contained in the $\alpha$
factor), one then arrives at the standard ST linewidth form, Eq.~(\ref{eq:STlw}).

However, the actual power emitted can be calculated using the true spatially-varying
fields using Poynting's theorem \cite{jackson}, as
\begin{equation}
P = \frac{I_0 \omega_0}{2\pi} \int_c \im[-\varepsilon] |\Psi_0(x)|^2 dx.
\end{equation}
Using specific form of the imaginary part of the dielectric function for a two-level gain medium 
(and using real units, not SALT units), the linewidth can then be written as
\begin{equation}
\delta \omega = \frac{\hbar \omega_0}{2 P} \int \gamma_{sp}(x) D(x) n_{sp}(x) dx \int \gamma_{sp}(x) D(x) dx. \label{eq:LW}
\end{equation}

Here we have derived a more physically intuitive and accurate form of the laser linewidth formula in which 
the critical role of spontaneous emission in the gain medium leaps out:~the linewidth is determined by the
spatially-varying, mode-dependent spontaneous emission rate and by the steady-state saturated inversion in
the presence of spatial hole-burning. Only if one makes the approximation of $\int \gamma_{sp} D dx \approx \gamma_c$ 
does this reduce to the standard Schawlow-Townes formula, and in making this substitution one tends to hide
the true origin of the linewidth, which is spontaneous emission.

While the above argument captures the basic physics of the ST formula in an improved form, it is not
a rigorous general result, as is the formula derived by Pick \textit{et al.}\ \cite{pick_linewidth_2015}, which includes
essentially all known corrections in a more accurate form.  However the N-SALT linewidth does have the same general features
as our simpler result, Eq.~(\ref{eq:LW}), and for a single lasing mode can be written as \cite{pick_linewidth_2015},
\begin{multline}
\delta \omega_{\textrm{N-SALT}} = \frac{\hbar \omega_0}{2 P} (1 + \tilde{\alpha}^2) \\
\times \frac{\omega_0^2\int \im[\varepsilon] |\boldsymbol{\Psi}_0|^2 d\mathbf{x} \int \im[\varepsilon]\frac{N_2(\mathbf{x})}{D(\mathbf{x})}|\boldsymbol{\Psi}_0|^2 d\mathbf{x}}
{\left|\int \boldsymbol{\Psi}_0^2 \left(\varepsilon + \frac{\omega_0}{2} \frac{d\varepsilon}{d\omega}|_{\omega_0} \right) d\mathbf{x} \right|^2}, \label{eq:NSALT}
\end{multline}
where $\boldsymbol{\Psi}_0$ is the normalized SALT lasing mode, and $\tilde{\alpha}$ is a generalization of the Lax-Henry $\alpha$ factor.
Very recently this formula has been quantitatively confirmed using first-principles numerical simulations \cite{cerjan_noise_2015}
for laser parameters where it predicts significant departures from the standard (corrected) form of the Schawlow-Townes formula.
Note that in Eq.~(\ref{eq:NSALT}) the cavity decay rate is completely absent, while the imaginary part
of the response of the gain medium plays a critical role. 

%The additional complexity found
%in Eq.~(\ref{eq:NSALT}) when compared against Eq.~(\ref{eq:LW}) stems from the inclusion of all known corrections
%to the basic form of the Schawlow-Townes linewidth, such as the incomplete inversion factor, which
%was also included in the approximate form of the linewidth through
%Eq.~(\ref{eq:incomp}) \cite{lax_conf_1966}. Both the Petermann factor and the Henry $\alpha$-factor
%were discovered during the development of semiconductor lasers. The Petermann factor corrects
%for the increase in the linewidth due to the non-Hermitian nature of the lasing modes, as discussed
%in Sec.~\ref{sec:Modes} \cite{petermann_calculated_1979,haus_excess_1985,siegman_excess_1989,hamel_nonorthogonality_1989,hamel_observation_1990}.
%The Henry $\alpha$-factor accounts for he coupling between intensity and phase fluctuations \cite{henry82,henry86}.
%Originally this was calculated as the ratio of the imaginary and real parts of the index of refraction
%of the gain medium \cite{henry82,henry86}, however eigenvector corrections have been found to this
%quantity \cite{pick_linewidth_2015,cerjan_noise_fdtd}. Finally, the bad-cavity correction is the only known correction that
%leads to a reduction in the predicted linewidth, and has been interpreted as linewidth narrowing due to an effective increase
%in the cavity $Q$ from the high dispersion of the gain medium, which reduces the group velocity
%of the light in the cavity \cite{lax_conf_1966,kolobov_role_1993,kuppens_quantum-limited_1994,vanexter_theory_1995,kuppens_evidence_1995}.
%
\section{Conclusion}

In summary we have presented a scattering framework for semiclassical laser theory and showed
how precisely it leads to zero laser linewidth, due to the discreteness of the poles of the scattering
matrix. This approach treats the full system as infinite from the beginning and does not introduce a
finite ``universe" and impose a limiting procedure and damping to recover the infinite system result,
although such an approach is a valid alternative method if the limiting absorption principle is observed.
We were inspired to analyze this question by the classic work of Lang \textit{et al.}\, which introduced
the notion of ``modes of the universe".  Our scattering point of view underlies recently developed \textit{ab initio} laser
theory (SALT), which leads to efficient computational tools for calculating steady-state lasing properties.
A generalization of the \textit{ab initio} theory to treat noise allows accurate calculation of 
the intrinsic linewidth due to quantum fluctuations when one goes beyond semiclassical laser theory.
We emphasize that the quantum-limited linewidth depends directly on the properties of the 
gain medium in the active lasing state, and it may be useful to write it in a form which does not
involve the passive cavity decay rate, but rather emphasizes the spontaneous emission rate.

\section*{Acknowledgements}
We acknowledge partial support by NSF grant DMR-1307632.  ADS thanks Marlan Scully
for drawing our attention to the work of Lang \textit{et al.}
We thank Steven Johnson and Adi Pick for very useful conversations about the laser
linewidth, noise effects, Perfectly Match Layers and the limiting absorption principle.
We would also like to thank Li Ge for stimulating discussions on methods for finding
the poles of the S-matrix for two-dimensional cavities.

%%%%%%%%%%%%%%%%%%%%%%%%%%%%%%%%
\appendix

\section{Review of SALT \label{app:SALT}}

In this Appendix we will provide a brief review of SALT to assist in understanding
the discussion in Sec.~\ref{sec:tlm}, and the results presented in Sec.~\ref{sec:example}.
However, the interested reader is referred to works dedicated to the SALT and the CF basis \cite{tureci06,tureci08,ge10},
and its elaborations to treat amplified modes \cite{cerjan_isalt_2014}, and complex gain
media \cite{cerjan12,cerjan_csalt_2015}.
The SALT equations find the steady-state solutions to the semiclassical
Maxwell-Bloch equations, for which the wave equation was given above in Eq.~(\ref{eq:maxWave}).  For this 
brief review we will only consider the simplest case of two-level atomic gain media, which leads to
the Bloch equations for the evolution of the atomic polarization:
inversion are
\begin{align}
\partial_t \mathbf{P}_g^+(\mathbf{x},t) =& -\left(\gamma_{\perp} + i \omega_{a}\right) \mathbf{P}_g^+ - \frac{i d}{\hbar}\left(\boldsymbol{\theta} \cdot \mathbf{E} \right) \boldsymbol{\theta}^*, \label{eq:MB1} \\
\partial_t D(\mathbf{x},t) =& -\gamma_\parallel(D - D_0(\mathbf{x})) - \frac{2}{i\hbar} \left(\mathbf{P}_g^- - \mathbf{P}_g^+ \right) \cdot \mathbf{E}. \label{eq:MB2}
\end{align}
%The motion of the poles of the scattering matrix of an active optical cavity
%subjected to an incident signal, studied in Sec.~\ref{sec:example}, depends upon the wavelength of that injected signal
%through the effects of gain saturation.
%Thus a proper treatment of the location of the resonances in an active optical
%cavity requires an understanding of the spatial degrees of freedom available.
%Furthermore, as discussed in Sec.~\ref{sec:Modes}, we expect the modes associated
%with the resonances of the cavity to be complex-valued, as they must satisfy
%the non-Hermitian Sommerfeld radiation condition. 
%In this section we review the
%steady-state \textit{ab initio} laser theory (SALT) and demonstrate how it solves
%the Maxwell-Bloch equations, Eqs.~(\ref{eq:MB1})--(\ref{eq:maxWave}) nearly exactly \cite{tureci06,tureci08,ge10},
%allowing it to calculate the behavior of the poles shown in Sec.~\ref{sec:example}.

For an electric field with a single frequency
component, or operating in the stable multi-mode regime, we can expand the electric
field over its $N$ constituent frequency components as
\begin{equation}
\mathbf{E}^+(\mathbf{x},t) = \sum_\sigma^{N} \mathbf{E}_\sigma(\mathbf{x})e^{-i\omega_\sigma t}, \label{eq:Eexp}
\end{equation}
where $\mathbf{E}^+$ is the positive frequency component of the electric field,
$\mathbf{E} = 2\textrm{Re}[\mathbf{E}^+]$.
In this expansion we make no distinction between lasing modes, which are associated with 
the poles of the scattering matrix, and amplified modes due to injected signals, which are externally generated. 

The critical assumption in SALT is the stationary inversion approximation (SIA), which states
that $\partial_t D = 0$,
even when multiple frequencies in the electric field are present; i.e. the populations do not respond to
beating terms between the modes, that would lead to four wave mixing and frequency comb generation.
This assumption requires that the non-radiative relaxation time of the gain medium
is much slower than these beating terms, $\gamma_\parallel \ll \Delta$, where $\Delta$
is the minimum frequency difference $\Delta = \omega_\sigma -\omega_\nu$ present
in the set of frequencies comprising the electric field, $\{\omega_\sigma\}$, and is satisfied by nearly all
microcavity lasers. In the parlance developed for the study of temporal chaotic
laser behavior, the SIA is valid for class A and B lasers, but not class C \cite{arecchi_ABC_84,ohtsubo}.
For class B lasers, which include most semiconductor based devices,
the mode spacing must also be well separated from the relaxation oscillation
frequency, so as to avoid resonantly driving relaxation fluctuations which could destabilize the
multimode solution. The relaxation oscillation frequency is $\omega_{\textrm{RO}} \sim \sqrt{\kappa \gamma_\perp}$ \cite{arecchi_ABC_84}, 
where $\kappa = \gamma_c/2$ is the field decay rate of the cavity. 
For microcavities without an injected signal, $\kappa \le \Delta_{\textrm{FSR}}$,
the mode spacing from the free spectral range of the cavity, so
$\omega_{\textrm{RO}} \le \sqrt{\Delta_{\textrm{FSR}} \gamma_\parallel} < \Delta_{\textrm{FSR}}$, and
as such, the SIA is still valid when $\gamma_\parallel < \Delta_{\textrm{FSR}}$ \cite{cerjan_isalt_2014}.
Furthermore, SALT has been rigorously tested
using direct finite-difference time-domain (FDTD) simulations of the Maxwell-Bloch
equations, with excellent quantitative agreement found \cite{ge08,cerjan12,cerjan_isalt_2014,cerjan_csalt_2015}.

As the SIA removes any time dependence from the inversion, the polarization must now
contain exactly the same frequency components as the electric field,
\begin{equation}
\mathbf{P}_g^+(\mathbf{x},t) = \sum_\sigma^{N} \mathbf{p}_\sigma(\mathbf{x})e^{-i\omega_\sigma t}.
\end{equation}
Using the multimode expansion for the electric field, Eq.~(\ref{eq:Eexp}), the polarization equation, Eq.~(\ref{eq:MB1}), 
can be solved independently for each frequency component, providing an expression for $\mathbf{p}_\sigma$ in terms of $D$ and $\mathbf{E}_\sigma$, 
and inserted into the wave equation as
\begin{multline}
\left[ \nabla \times \nabla \times - \left(\varepsilon_c(\mathbf{x}) + \frac{4\pi |\theta|^2 D(\mathbf{x})}{\hbar(\omega_\sigma - \omega_a + i \gamma_\perp)}\right)k_\sigma^2 \right]\mathbf{E}_\sigma(\mathbf{x}) \\ = 0, \label{eq:SALT1}
\end{multline}
which explicitly identifies the gain due to the inversion of the atomic medium,
and in which $\theta$ has been taken to be collinear to the electric field.
Likewise, using Eq.~(\ref{eq:MB2}), we can solve directly for the inversion as
\begin{equation}
D(\mathbf{x}) = \frac{D_0(\mathbf{x})}{1 + \frac{4 |\theta|^2}{\hbar^2 \gamma_\perp \gamma_\parallel} \sum_\sigma^N \Gamma_\sigma |\mathbf{E}_\sigma|^2}, \label{eq:SALT2}
\end{equation}
in which the denominator contains the effects of gain-saturation due to the electric
field within the cavity, and $\Gamma_\sigma = \gamma_\perp^2 / ( (\omega_\sigma -\omega_a)^2 + \gamma_\perp^2)$ is
the Lorentzian gain curve. Together, Eqs.~(\ref{eq:SALT1}) and (\ref{eq:SALT2})
form the fundamental SALT equations (for the two level gain case): $N$ self-consistent, differential equations for the
unknown spatial profiles of the $N$ modes in the electric field at N unknown frequencies, 
coupled through the non-linear spatial hole-burning in the inversion. These are to be solved with outgoing
Sommerfeld boundary conditions at infinity, which can only be satisfied at discrete frequencies. The SALT
equations also reveal the natural units for the electric field and atomic inversion,
$E_{\textrm{SALT}} = 2|\theta|/(\hbar \sqrt{\gamma_\perp \gamma_\parallel}) E$, and
$D_{\textrm{SALT}} = (4\pi |\theta|^2/\hbar \gamma_\perp) D$. 

The numerical
solution of the SALT equations is performed by expanding the mode profile functions
over a basis,
\begin{equation}
\mathbf{E}_\sigma(\mathbf{x}) = \sum_n a_n^{(\sigma)} \mathbf{f}_n(\mathbf{x};\omega_\sigma),
\end{equation}
and then using a non-linear solver to find the expansion coefficients $a_n^{(\sigma)}$.
There are two different basis sets that have been used to solve the SALT equations.
The first, and more developed, is based on a set of constant flux (CF) states \cite{tureci06,ge10},
which are defined as
\begin{equation}
  \left[ -\nabla \times \nabla \times + \left(\varepsilon_c(\mathbf{x}) + \eta_n F(\mathbf{x}) \right) \mathbf{k}^2 \right] 
  \mathbf{u}_n(\mathbf{x}; \omega) = 0, \label{eq:CFgen}
\end{equation}
and satisfy the outgoing boundary condition at frequency $\omega$ at the surface of last scattering.
Here, $\eta_n(\omega)$ and $\mathbf{u}_n(\mathbf{x};\omega)$ are the eigenvalue and eigenstates of the CF equation,
and $D_0(\bx) = D_0 F(\bx)$ is the profile of the pump. A key aspect of the SALT approach is that for any given
frequency (wavevector) the CF states form a complete biorthogonal basis for an arbitrary purely outgoing solution;
hence they can expand any solution to the SALT equations.

Moreover, by observing the similarities between Eq.~(\ref{eq:CFgen})
and Eq.~(\ref{eq:SALT1}), it is clear that just above the first lasing threshold where the spatial hole-burning
in the inversion is negligible, the threshold lasing mode is given by a single CF basis state at the lasing
frequency satisfying
\begin{equation}
\eta(\omega) = \frac{\gamma_\perp D_0}{\omega - \omega_a + i \gamma_\perp}.
\end{equation}
Hence above threshold, the CF basis represents a numerically efficient basis for solving
the SALT equations, as each lasing mode can be represented by only a few CF states.
The Sommerfeld condition is implemented by matching conditions on the superposition at
the last scattering surface to purely outgoing free hermitian solutions.
The CF equation also facilitates solving the SALT equations for amplified modes by yielding
an additional basis, $\mathbf{v}_m(\mathbf{x};\omega)$, $\beta_m(\omega)$, which are also
defined by Eq.~(\ref{eq:CFgen}), but satisfy a purely incoming boundary condition instead,
thus representing a signal with a constant flux into the cavity.

Recently an alternative approach for solving the SALT equations has been developed, 
using a position
space basis, and a truncated simulation region bordered by a perfectly matched layer \cite{esterhazy14}.
This follows the conceptual approach discussed in detail at the end of Sec.~\ref{sec:Modes}. The Sommerfeld
condition is not imposed by matching at a last scattering surface, but rather by the presence of the PML.
In general, the CF methods allow for more elegant theoretical analysis, while the position space basis
is ultimately expected to be more efficient computationally. In both approaches a non-linear iteration scheme
will need to be implemented to find the correct self-consistent solution in the basis of choice.

The SALT modal wave equation, Eq.~(\ref{eq:SALT1}) is satisfied by both lasing modes and injected signals,
which interact and compete with each other for gain through the non-linear spatial hole-burning denominator
of Eq.~(\ref{eq:SALT2}). There
are two major differences between lasing modes and injected signals though; the
former must satisfy the Sommerfeld radiation condition, while the latter satisfies
a mixed boundary condition with both incoming and outgoing terms. As a direct consequence
of this, the frequency of a lasing mode needs to be determined self-consistently from the outgoing
boundary condition itself,
while the frequency of an injected signal is externally fixed, and the mixed boundary
condition can be satisfied at any frequency. 

\section{Partially reflecting mirrors in SALT \label{app:B}}

Traditionally, SALT has been used to simulate semiconductor based microlasers, in which
the optical confinement for one-dimensional systems is provided by Fresnel reflection due
to the dielectric mismatch across the cleaved facets of the device. However, for even the
materials with the largest index of refraction, the optical confinement due to Fresnel
reflection is weak, $R < .36$ for $n < 4$. Thus to simulate high finesse systems, one must
either introduce a geometric solution, such as a Bragg reflector, or a partially
reflecting mirror. In the results presented in Sec.~\ref{sec:example} we chose to use
a partially reflecting mirror, as we have the
additional motivation of simulating a similar system as described by Lang \textit{et al.}\ and others \cite{lang_why_1973,baseia_semiclassical_1984,penaforte_quantum_1984,dutra_spontaneous_1996}.
In this appendix we briefly detail how to implement a partially reflecting mirror in SALT
simulations.

\begin{figure}[t!]
\centering
\includegraphics[width=0.45\textwidth]{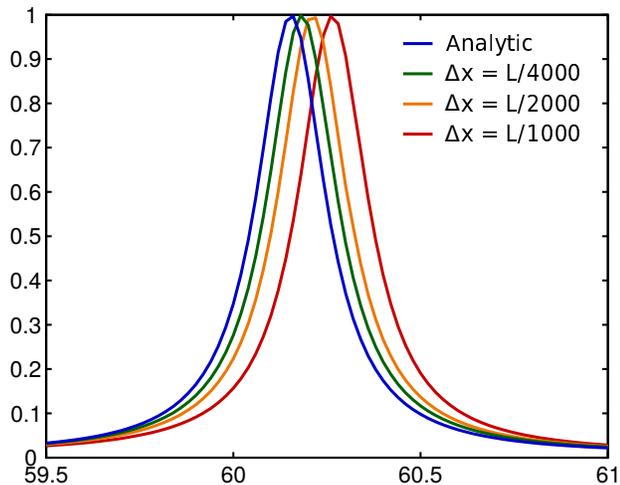}
\caption{Plot of the transmitted intensity of an injected signal with frequency $\omega_{\IN}$, 
into the passive, two-sided cavity with partially reflecting mirrors ($R=0.8$) on both ends studied
in Fig.~\ref{fig:passive}.
Different colors represent different choices of resolution, $\Delta x$ (red, orange, and green),
or the analytic solution (blue).
\label{fig:dx}}
\end{figure}

Mathematically, a partially reflecting mirror can be described by a delta function in the
passive cavity dielectric function,
\begin{equation}
\varepsilon_{c,total}(x) = \varepsilon_c(x) + \sum_m \Lambda_m \delta (x-x_m),
\end{equation}
where $\Lambda_m$ represents the strength of the $m$th mirror at $x_m$. For a single
mirror, it is straightforward to show that the reflection and transmission coefficients
are
\begin{align}
R = \frac{\left(\frac{k\Lambda}{2}\right)^2}{1 + \left(\frac{k\Lambda}{2}\right)^2}, \\
T = \frac{1}{1 + \left(\frac{k\Lambda}{2}\right)^2},
\end{align}
which demonstrates that the figure of merit for the reflectivity of a partially reflecting
mirror is $k \Lambda$.
There are two different ways to implement this delta-function in SALT, either by incorporating
the partially reflecting mirrors directly into the passive cavity dielectric, as indicated
above, or for mirrors at the cavity boundary, adjusting the outgoing boundary condition \cite{cerjan_thesis}.
The former method is numerically more stable, and is implemented in a discretized
system as
\begin{equation}
\varepsilon_{c,total}(x_n) = \varepsilon_c(x_n) + \sum_m \frac{\Lambda_m}{\Delta x} \delta_{x_n,x_m}, \label{eq:pixel}
\end{equation}
where $\delta_{ij}$ is the Kronecker delta, $x_n$ is the spatial position of the $n$th pixel
in the system, and $\Delta x$ is the resolution. Despite the fact that the reflectivity
of such a mirror is dependent upon $\zeta = k\Lambda$, when using the CF basis to solve
the SALT equations, one can
choose $\zeta$ to be a constant, and then adjust $\Lambda$ accordingly to achieve a
system with reflectivity independent of the frequency, something which is not possible in
time-domain simulation techniques.

We can confirm that the discretized representation of a partially reflecting mirror used in
conjunction with the SALT algorithm is correct by comparing it against the analytic result.
In Fig.~\ref{fig:dx} we show the transmission resonance for the same passive cavity shown
in Fig.~\ref{fig:passive} when calculated analytically (blue), and simulated using
I-SALT with increasing resolution (red, orange, and green). The effect of discretizing
the system is to shift the cavity resonances slightly, and as the resolution is increased,
the location of the cavity resonance converges to the analytic value, as shown. However,
regardless of the spatial resolution, the Lorentzian shape of the cavity resonance remains unchanged.

\end{document}